\pdfoutput=1
\documentclass[aip,jcp,preprint,amsmath,amssymb]{revtex4-2}

\usepackage{graphicx}
\usepackage{bm}
\usepackage[T1,OT1]{fontenc}
\DeclareTextCommand{\k}{OT1}[1]{{\fontencoding{T1}\fontfamily{lmr}\selectfont\k{#1}}}

\newcommand{\nc}{\ensuremath{n_{\mathrm{c}}}}
\newcommand{\gtc}{\ensuremath{g^{(2)}_{\mathrm{c}}}}
\newcommand{\lamd}{\ensuremath{\boldsymbol{\lambda}\cdot\hat{\mathbf{d}}}}
\newcommand{\rsq}{\ensuremath{s}}
\newcommand{\Ebar}{\ensuremath{\bar{E}}}
\newcommand{\Var}{\ensuremath{\mathrm{Var}}}
\newcommand{\dE}{\ensuremath{\Delta E}}
\newcommand{\dP}{\ensuremath{\delta_{P}}}

\begin{document}

\title{Vacuum-Field Squeezing by Molecules in Optical Cavities: An Exact
Bound and Linear-Response Theory}

\author{Kurt R. Brorsen}
\email{brorsenk@missouri.edu}
\affiliation{Department of Chemistry, University of Missouri, Columbia, Missouri 65211, United States}

\begin{abstract}
The ground state of a molecule coupled to an optical cavity contains
virtual photons, and squeezing of the photonic field has been found in
exactly solvable models, perturbation theory, and numerical calculations.
Here, two exact sum rules for the field quadratures show that, within
the dipole approximation, the vector-potential fluctuations of a linearly
polarized mode never exceed their vacuum value in the ground state of any
molecule at any coupling strength. Consequently,
the field is squeezed whenever the mode couples to a dipole-allowed
transition from a nondegenerate ground state, and the cavity-induced
ground-state energy shift cannot increase with the cavity frequency at a
fixed coupling vector, which is an exact constraint on approximate methods.
Moreover, the incoherent photon statistics, which exclude the
coherent part of the field, are never sub-Poissonian.
At weak coupling, an exact resolvent expression reduces the squeezing and
the ratio of the anomalous to the normal photon moment to linear-response
properties of the isolated molecule, with the ratio set by a mean
excitation energy. Time-dependent density functional theory with CAM-B3LYP
and PBE0 reproduces exact and QED coupled-cluster references for both
quantities to within 5\% at 45 benchmark points. A survey of 29 molecules
gives mean excitation energies of 9.5--21~eV at a cavity frequency of 2~eV
and shows that the most polarizable molecules squeeze the field the most.
\end{abstract}

\maketitle

\section{Introduction}
\label{sec:intro}

In polaritonic chemistry, strong coupling between molecular transitions
and the quantized field of an optical cavity modifies chemical
properties and reactivity.\cite{Ebbesen2016,Herrera2020,GarciaVidal2021,Fregoni2022}
\textit{Ab initio} cavity quantum electrodynamics (QED) methods, including QED
Hartree--Fock (QED-HF), QED coupled-cluster (QED-CC), and QED
density-functional theory, treat the molecule and the cavity mode on the
same footing and are now routinely used to investigate how a cavity modifies
molecular ground and excited
states.\cite{Flick2017,Haugland2020,Mordovina2020,DePrince2021,Riso2022,Pavosevic2023,Foley2023,Ruggenthaler2023,Bauman2026}
In these approaches, the ground state of the coupled system contains
virtual photons, and its photonic part carries information about
electron--photon correlation that the molecular energy alone does not
reveal.

Correlated \textit{ab initio} studies have
reported mean photon numbers\cite{Weber2025,Hassan2024} and the state of the photon
field.\cite{Tang2025}
Additionally, intracavity photon
statistics computed from QED-CC wave functions that were validated against exact
diagonalization were recently reported.\cite{Brorsen2026}
In contrast to these correlated
treatments, the photon subsystem of the QED-HF reference is a coherent
state.\cite{Haugland2020} In the strong-coupling QED-HF
reference,\cite{Riso2022} each molecular orbital carries its own coherent
state, and the photon subsystem is a statistical mixture of coherent
states whose light is classical.\cite{ElMoutaoukal2026} Nonclassical properties of
the cavity field, such as squeezing of a field quadrature below its vacuum
value,\cite{Walls1983,LoudonKnight1987} therefore lie beyond a
coherent-state description.

Squeezing of the vacuum field by matter has been studied many times. Persico and
Power, Passante and Power, and Power and Thirunamachandran used
second-order perturbation theory to show that the electric energy density of the
vacuum field is raised and the magnetic energy density is lowered near a
ground-state atom or
molecule.\cite{PersicoPower1986,PassantePower1987,PowerThiru1992} Drummond
exactly solved a harmonic-oscillator photodetector coupled to a single cavity mode
to show that the fluctuations of the magnetic field are reduced
and those of the electric field are increased.\cite{Drummond1987} Becker
et al.\ had analogous findings for a free charge in a cavity and suggested that
this squeezing would be present in the vicinity of atoms and molecules
in general.\cite{Becker1987} Similarly, Glauber and Lewenstein showed
that, in the ground state of a cavity filled with a dielectric, the magnetic
field fluctuates less and the displacement field more than in the
vacuum.\cite{GlauberLewenstein1991}

In the ultrastrong and deep strong coupling regimes of cavity
QED,\cite{FriskKockum2019,FornDiaz2019} the ground states of the
Hopfield\cite{Hopfield1958} and quantum Rabi models are
squeezed,\cite{Emary2003,Ciuti2005,Ashhab2010,Hwang2015,Stassi2016,DeBernardisJaako2018,Larson2024,KamHu2026} although this squeezing
is not transferred to the field emitted by the cavity.\cite{Stassi2016}
For molecules, Triana and Herrera found weak ground-state squeezing of a
cavity mode coupled to a single molecular vibration described by a Morse
potential, and stronger, transient squeezing under ultrafast modulation of
the cavity frequency.\cite{Triana2022} Tasci et al.\ obtained squeezed and
super-Poissonian ground-state light for argon chains\cite{Tasci2025} from
the photon many-body dispersion functional,\cite{TasciPRL2025} which maps
the coupled system onto an effective quadratic Hamiltonian. Moreover,
Tang et al.\ computed the ground state of two H$_2$ molecules in a cavity
with a deep-learning variational Monte Carlo method and found the photonic
Wigner function squeezed along the quadrature proportional to the vector
potential.\cite{Tang2025} They explained this squeezing qualitatively with
a second-order effective Hamiltonian for the photon field, whose ground
state is a squeezed vacuum. However,
a proof for arbitrary matter at arbitrary coupling, a systematic determination of the size of the
squeezing across molecules, and a quantitative way to predict it without a
correlated QED calculation have been lacking.

Sum rules have long provided constraints on the ground states of
cavity-coupled systems. The Thomas--Reiche--Kuhn (TRK) sum rule forbids a superradiant
phase transition of atoms coupled to a cavity mode through electric-dipole
transitions,\cite{Rzazewski1975,Nataf2010} and gauge invariance excludes
this transition for any atoms or molecules in which only the charges, not
intrinsic magnetic moments, couple to a finite number of modes in the
dipole approximation.\cite{BialynickiBirula1979} Similarly,
the cancellation between paramagnetic and diamagnetic responses enforced
by the TRK sum rule prevents photon condensation in a spatially uniform
cavity field.\cite{Andolina2019}
Barnett and Loudon showed that the commutation relations of the field
operators in a magnetodielectric medium imply sum rules for the phase and
group velocities of its polaritons.\cite{BarnettLoudon2012}
More recently, Savasta et al.\ derived a TRK sum rule for the cavity field
itself that holds for arbitrary matter coupled to a field
coordinate.\cite{Savasta2021} However, to our
knowledge, sum rules of this type have not been used to constrain the
quantum fluctuations of the cavity field in the ground state.

In this work, we show that two exact sum rules for the field quadratures,
combined with the Cauchy--Schwarz inequality, bound the fluctuations of the
vector potential of a linearly polarized cavity mode in the ground state of
any molecule at any coupling strength within the dipole approximation. This result establishes, for a
single cavity mode, the general squeezing anticipated by Becker et
al.\cite{Becker1987} As a consequence, the cavity field is squeezed
whenever the cavity mode couples to a dipole-allowed transition from a
nondegenerate ground state, and the incoherent photon statistics are never
sub-Poissonian. Although the
ground-state squeezing is not emitted by the cavity,\cite{Stassi2016} its
amount equals the rate at which the cavity-induced shift of the exact
ground-state energy decreases with the cavity frequency at a fixed coupling
vector.
This shift therefore never increases with the frequency, which gives an
exact constraint for approximate methods in polaritonic chemistry.
We also show that, for a nondegenerate
ground state, the second moments of the field are fixed by the spectral
weights of the field coordinate in the coupled system, with all poles
included.

We then derive an exact expression for the photon moments that reduces at
weak coupling to the linear response of the isolated molecule, which we
benchmark against exact diagonalization and QED-CCSD-22 at 45 points and
apply to 29 diverse molecules.

\section{Theory}
\label{sec:theory}

\subsection{Hamiltonian and photon moments}
\label{sec:moments}

For a molecule coupled to a single cavity mode within the dipole
approximation, the Pauli--Fierz (PF) Hamiltonian in the length
gauge\cite{Rokaj2018,Schafer2020,Haugland2020} is
\begin{equation}
  \hat{H} \;=\; \hat{H}_{\mathrm{el}}
  \;+\; \omega\, \hat{b}^{\dagger}\hat{b}
  \;-\; \sqrt{\tfrac{\omega}{2}}\,(\lamd)\,(\hat{b}^{\dagger}+\hat{b})
  \;+\; \tfrac{1}{2}(\lamd)^{2},
  \label{eq:pf}
\end{equation}
where $\hat{H}_{\mathrm{el}}$ is the electronic Hamiltonian,
$\hat{\mathbf{d}}$ is the molecular dipole operator, and $\omega$ and
$\boldsymbol{\lambda}$ are the cavity frequency and coupling vector,
respectively. The final term is the dipole self-energy (DSE). We write
$\boldsymbol{\lambda} = \lambda\mathbf{e}$ with the real polarization unit
vector $\mathbf{e}$ of a linearly polarized mode and denote the dipole
component along the polarization by $\hat{d} = \mathbf{e}\cdot\hat{\mathbf{d}}$. The field quadratures are
$\hat{X} = (\hat{b}+\hat{b}^{\dagger})/\sqrt{2}$ and
$\hat{P} = i(\hat{b}^{\dagger}-\hat{b})/\sqrt{2}$, with
$[\hat{X},\hat{P}] = i$ and vacuum variances of $1/2$. In the length gauge,
$\hat{X}$ is proportional to the transverse displacement field and
$\hat{P}$ to the vector potential, and hence to the magnetic field of the
mode.\cite{CohenTannoudji1989,Schafer2020} Because the
Power--Zienau--Woolley transformation that connects the length gauge to the
Coulomb gauge commutes with the vector potential,\cite{Schafer2020}
$\hat{P}$ represents the same field observable in both gauges when the
molecular basis is complete.

A shift of the coordinate origin or of the dipole convention changes the
exact ground state of Eq.~\eqref{eq:pf} only by a coherent displacement of
the photon mode,\cite{Schafer2020,Brorsen2026} so the statistics of the field are
defined through the fluctuation operator
$\delta\hat{b} = \hat{b} - \langle\hat{b}\rangle$. In terms of the
incoherent photon number $\nc = \langle\delta\hat{b}^{\dagger}\delta\hat{b}\rangle$
and the anomalous moment $m = \langle\delta\hat{b}^{2}\rangle$, the
quadrature variances are
\begin{equation}
  \Var(X) = \tfrac{1}{2} + \nc + \mathrm{Re}\,m, \qquad
  \Var(P) = \tfrac{1}{2} + \nc - \mathrm{Re}\,m .
  \label{eq:quadvar}
\end{equation}
The smallest variance among all quadratures, $\tfrac{1}{2} + \nc - |m|$,
falls below the vacuum value exactly when $|m| > \nc$, which is the
condition for principal squeezing.\cite{Luks1988} Because squeezing is a
property of a choice of field variables rather than of the state
alone,\cite{GlauberLewenstein1991} we measure it, as is usual for
intracavity fields, against the vacuum of the bare cavity mode. We
characterize the squeezing of the field by its amount,
$\dP = \tfrac{1}{2} - \Var(P) = \mathrm{Re}\,m - \nc$, and by the ratio
\begin{equation}
  \rsq = \frac{|\langle\delta\hat{b}^{2}\rangle|}{\langle\delta\hat{b}^{\dagger}\delta\hat{b}\rangle}
       = \frac{|m|}{\nc},
  \label{eq:r}
\end{equation}
which exceeds unity for a squeezed field.
For any state with a nonnegative
Glauber--Sudarshan $P$ function,\cite{Glauber1963,Sudarshan1963} which
includes coherent states and their statistical mixtures, $|m| \le \nc$,
so that $\rsq \le 1$ whenever $\nc > 0$. The QED-HF and strong-coupling QED-HF references
belong to this class, so $\rsq > 1$ signals electron--photon correlation
beyond these coherent-state references. Finally, the degree of second-order coherence
of the field fluctuations is
$\gtc = \langle\delta\hat{b}^{\dagger 2}\delta\hat{b}^{2}\rangle/\nc^{2}$,\cite{Brorsen2026}
and the Cauchy--Schwarz inequality
$\langle\delta\hat{b}^{\dagger 2}\delta\hat{b}^{2}\rangle \ge |\langle\delta\hat{b}^{2}\rangle|^{2}$
gives $\gtc \ge \rsq^{2}$ for any state.

\subsection{Exact bound on the vector-potential fluctuations}
\label{sec:bound}

Adding $\omega/2$ to Eq.~\eqref{eq:pf} and expressing the photon operators
through the quadratures gives
\begin{equation}
  \hat{H} + \frac{\omega}{2} \;=\; \hat{H}_{\mathrm{el}}
  \;+\; \frac{\omega}{2}\,\hat{P}^{2}
  \;+\; \frac{1}{2}\left(\sqrt{\omega}\,\hat{X} - \lambda\hat{d}\right)^{2},
  \label{eq:square}
\end{equation}
in which the DSE completes the square. Thus, the molecule couples to the
field only through $\hat{X}$, and the only term that does not commute with
$\hat{X}$ is $\omega\hat{P}^{2}/2$. The double commutators of the
quadratures with $\hat{H}$ are therefore c-numbers for any electronic
Hamiltonian and any dipole operator,
\begin{equation}
  [\hat{X},[\hat{H},\hat{X}]] = \omega, \qquad
  [\hat{P},[\hat{H},\hat{P}]] = \omega .
  \label{eq:dcomm}
\end{equation}
We assume that $\hat{H}$ has a normalizable ground state $|\Psi_0\rangle$
in which the moments of $\hat{X}$, $\hat{P}$, and $\hat{d}$ used below are
finite, and sums over states include integrals over any continuum.
Taking the ground-state expectation value of Eq.~\eqref{eq:dcomm} and
inserting the exact eigenstates $|\Psi_n\rangle$ of $\hat{H}$, with
excitation energies $\dE_n = E_n - E_0$, gives the energy-weighted sum
rules
\begin{equation}
  \sum_{n} \dE_n\,|X_{n0}|^{2} = \frac{\omega}{2}, \qquad
  \sum_{n} \dE_n\,|P_{n0}|^{2} = \frac{\omega}{2},
  \label{eq:sumrules}
\end{equation}
where $X_{n0} = \langle\Psi_n|\hat{X}|\Psi_0\rangle$ and $P_{n0}$ is defined
analogously. The first sum rule is the field TRK sum rule of Savasta et
al.\cite{Savasta2021} for $\hat{X}$, which is the only field coordinate
that couples to the matter in the length gauge. For a complete matter
basis, the second sum rule is the same field sum rule written for the
vector potential in the Coulomb gauge. The commutator
$[\hat{H},\hat{X}] = -i\omega\hat{P}$ relates the two sets of matrix
elements, $P_{n0} = i\,\dE_n X_{n0}/\omega$,\cite{Savasta2021} so that the
second sum rule becomes
\begin{equation}
  \sum_{n} \dE_n^{3}\,|X_{n0}|^{2} = \frac{\omega^{3}}{2} .
  \label{eq:s3}
\end{equation}
Because $\langle\Psi_0|[\hat{H},\hat{X}]|\Psi_0\rangle = 0$, the
expectation value of $\hat{P}$ vanishes and
$\Var(P) = \sum_n |P_{n0}|^{2} = \omega^{-2}\sum_n \dE_n^{2}|X_{n0}|^{2}$.
The Cauchy--Schwarz inequality
\begin{equation}
  \Big(\sum_n \dE_n^{2}|X_{n0}|^{2}\Big)^{2} \;\le\;
  \Big(\sum_n \dE_n|X_{n0}|^{2}\Big)\Big(\sum_n \dE_n^{3}|X_{n0}|^{2}\Big),
  \label{eq:cs}
\end{equation}
which is the moment inequality familiar from nuclear sum
rules,\cite{Bohigas1979,Lipparini1989} then yields
\begin{equation}
  \Var(P) \;\le\; \frac{1}{2} .
  \label{eq:bound}
\end{equation}

For a nondegenerate ground state and $\lambda \neq 0$, equality holds if
and only if the ground state is an eigenstate of $\hat{d}$, in which case
the photon mode is a displaced vacuum and is decoupled from the molecule
(supplementary material). The
vector-potential quadrature of a linearly polarized mode is therefore never
anti-squeezed in the ground state, and it is strictly squeezed whenever the
cavity mode couples to a dipole-allowed transition from a nondegenerate
ground state (in a complete basis; supplementary material).
Equation~\eqref{eq:bound} is the general, nonperturbative
form of the squeezing found in the exactly solvable models of
Refs.~\citenum{Drummond1987} and \citenum{GlauberLewenstein1991} and, as a
limiting case, in the free-charge model of Ref.~\citenum{Becker1987}. Moreover,
with the uncertainty relation $\Var(X)\Var(P) \ge 1/4$, it gives
$\Var(X) \ge 1/2$, so the displacement quadrature is never squeezed, in
agreement with the rigorous results for the Nelson and spin-boson
models.\cite{Betz2002,Hirokawa2014}

The derivation uses only two properties of Eq.~\eqref{eq:square}, namely
that the molecule couples to the field linearly through $\hat{X}$ and that
the field enters quadratically. Equation~\eqref{eq:bound} therefore holds
for any electronic Hamiltonian, including $\hat{H}_{\mathrm{el}}$ in a
truncated basis and with any one-electron form of the DSE, and for any
number of molecules. It also holds when the DSE is omitted, provided that
the Hamiltonian then has a ground state, because the double commutators of
Eq.~\eqref{eq:dcomm} are unchanged. For a multimode cavity, the same sum
rules written for the matrix of moments show that the covariance matrix of
the vector-potential quadratures of all modes is bounded by its vacuum
value, so that no combination of modes is anti-squeezed (supplementary
material).
For real mode functions, the magnetic field at any point is a linear
combination of these quadratures, so that the ground-state magnetic energy
density nowhere exceeds its vacuum value for any finite number of modes. This
result is the nonperturbative counterpart of the lowered magnetic energy
density found at second order near ground-state atoms and
molecules.\cite{PersicoPower1986,PassantePower1987,PowerThiru1992,Passante2018,PassanteRizzuto2025}

For a single circularly polarized mode, as in the chiral-cavity models of
Kulkarni et al.\ and El-Amrani et al.,\cite{Kulkarni2026,ElAmrani2026} the
molecule couples to both quadratures, so that Eq.~\eqref{eq:bound} need not
hold (supplementary material).
In contrast, El Moutaoukal et al.\ described molecules in a chiral cavity
with two counter-propagating modes whose circular polarization vectors have
opposite handedness, beyond the dipole approximation.\cite{ElMoutaoukal2026chiral}
Within the dipole approximation, such a pair is equivalent to two linearly
polarized modes, to which the multimode bound applies.
The proof also requires the dipole approximation, beyond which a
spatially varying cavity field can condense in electron systems with a
sufficiently large paramagnetic orbital susceptibility.\cite{Andolina2020}
In a truncated matter basis, the standard Hamiltonians of different gauges
are no longer equivalent,\cite{DeBernardis2018,DiStefano2019} and
Eq.~\eqref{eq:bound} is then a statement about the length-gauge
(dipole-gauge) Hamiltonian (supplementary material).

Combined with Eq.~\eqref{eq:quadvar}, Eq.~\eqref{eq:bound} gives
$\nc - \mathrm{Re}\,m \le 0$, and therefore $|m| \ge \mathrm{Re}\,m \ge \nc$
and
\begin{equation}
  \rsq \;\ge\; 1
  \label{eq:rge1}
\end{equation}
for every ground state of Eq.~\eqref{eq:pf} with $\nc > 0$. Moreover, because
$\gtc \ge \rsq^{2}$, it follows that $\gtc \ge 1$, so the incoherent photon
statistics of a cavity ground state are never sub-Poissonian. Thus, the
nonclassicality of the ground-state field is revealed by its quadrature
variances rather than by its second-order photon statistics, as El
Moutaoukal et al.\ noted for a squeezed vacuum.\cite{ElMoutaoukal2026}
For a complete matter basis, $\rsq \ge 1$ and $\gtc \ge 1$ also hold for the
photon operators of the Coulomb gauge, because the vector-potential
quadrature is the same observable in both gauges (Sec.~\ref{sec:moments}),
although the values of $\rsq$ and $\gtc$ differ between the gauges.\cite{StokesNazir2022}

Castagnola et al.\ wrote the derivative of the ground-state energy with
respect to the cavity frequency at a fixed coupling vector as
$\langle\hat{b}^{\dagger}\hat{b}\rangle$ plus $1/(2\omega)$ times the
expectation value of the bilinear coupling term.\cite{Castagnola2024wires}
The Hellmann--Feynman
theorem applied to Eq.~\eqref{eq:square}, combined with the virial relation
$\omega\langle\hat{P}^{2}\rangle =
\sqrt{\omega}\,\langle\hat{X}(\sqrt{\omega}\hat{X} - \lambda\hat{d})\rangle$
that follows from $\langle[\hat{X}\hat{P},\hat{H}]\rangle = 0$,\cite{Theophilou2020} gives
\begin{equation}
  \left(\frac{\partial E_0}{\partial\omega}\right)_{\boldsymbol{\lambda}}
  \;=\; \Var(P) - \frac{1}{2} \;\le\; 0,
  \label{eq:dEdw}
\end{equation}
where $E_0$ is the ground-state energy of Eq.~\eqref{eq:pf}. This energy
does not contain the zero-point energy $\omega/2$ of the cavity mode, so
that $E_0 - E_{\mathrm{el}}$, with $E_{\mathrm{el}}$ the ground-state
energy of the isolated molecule, is the shift of the ground-state energy
induced by the cavity. This shift is therefore a non-increasing function of the
cavity frequency at a fixed coupling vector, and its frequency derivative
equals $-\dP$.

The sum rules also express the second moments of the field through the
spectrum of $\hat{X}$. With the normalized weights
$w_n = 2\dE_n|X_{n0}|^{2}/\omega$, for which $\sum_n w_n = 1$ and
$\sum_n w_n\dE_n^{2} = \omega^{2}$ by Eqs.~\eqref{eq:sumrules} and
\eqref{eq:s3}, the quadrature variances of a nondegenerate ground state are
\begin{equation}
  \Var(P) = \frac{\langle\dE\rangle_{w}}{2\omega}, \qquad
  \Var(X) = \frac{\omega}{2}\,\langle\dE^{-1}\rangle_{w},
  \label{eq:spectral}
\end{equation}
where $\langle\cdot\rangle_{w}$ denotes the average over the weights $w_n$.
Because $\hat{X}$ is proportional to the displacement field, the weights
$w_n$ form the energy-weighted spectral function of this field,\cite{Flick2019} and
Eq.~\eqref{eq:spectral} expresses the ground-state squeezing through all
poles and residues of the response function of $\hat{X}$ (supplementary material).
Related exact
expressions in terms of response functions are known for the ground-state
photon number in linear dissipative media\cite{DeLiberato2017} and for the
bath fluctuations of the spin-boson model.\cite{BlundenCodd2017}

When the cavity spectrum consists of exactly two poles, a lower and an
upper polariton at $\omega_{-} < \omega < \omega_{+}$, the two sum rules fix
both weights, and Eq.~\eqref{eq:spectral} becomes
\begin{equation}
  \Var(P) = \frac{\omega^{2} + \omega_{+}\omega_{-}}{2\omega\,(\omega_{+}+\omega_{-})},
  \qquad
  \frac{1}{2} - \Var(P) =
  \frac{(\omega_{+}-\omega)(\omega-\omega_{-})}{2\omega\,(\omega_{+}+\omega_{-})},
  \label{eq:twopol}
\end{equation}
which is exact for the Hopfield model with a single bosonic matter mode.
For this model, Eq.~\eqref{eq:twopol} is equivalent to the exact
quadrature variance of Drummond,\cite{Drummond1987} written in terms of the
polariton frequencies.
In the limits $\omega_{-} \to 0$ and $\omega_{+} \to \infty$,
Eq.~\eqref{eq:twopol} reproduces the results of Becker et al.\ for a free
charge\cite{Becker1987} and of Glauber and Lewenstein for a
dielectric-filled cavity,\cite{GlauberLewenstein1991} respectively. Artoni
and Birman showed that the vacuum, number, and coherent states of exciton
polaritons are squeezed with respect to the corresponding states of a mixed
photon--exciton boson, by an amount that depends on the polariton
dispersion.\cite{ArtoniBirman1991} Equation~\eqref{eq:twopol} instead
refers to the vacuum of the bare cavity mode.
For two coupled
harmonic oscillators, Gietka\cite{Gietka2024} and Mirkhalaf et
al.\cite{Mirkhalaf2026} obtained the squeezing parameters of the ground
state from the frequency shifts of the normal modes. The sum rules extend
this relation to any coupled system in which $\hat{X}$ connects the ground
state to only two excited states, as for a single harmonic matter mode, and
Eq.~\eqref{eq:twopol} then expresses the squeezing of the cavity vacuum
through the lower-polariton, upper-polariton, and bare cavity frequencies
alone.
For other matter, such as ensembles of molecules with many electronic
excitations,\cite{Weight2026} Eq.~\eqref{eq:twopol} evaluated
with the two dominant poles is an approximation, and weak poles far above
the polaritons can still contribute to its error (supplementary material).

\subsection{Weak coupling and the linear-response route}
\label{sec:weak}

The photon moments can also be written exactly in terms of the dipole
fluctuations of the coupled system. With $\hat{K} = \hat{H} - E_0$ and
$\delta\hat{d} = \hat{d} - \langle\hat{d}\rangle$, the pull-through
relation\cite{Griesemer2001} for $\delta\hat{b}|\Psi_0\rangle$ and the
stationarity of the ground state give
\begin{equation}
  m = \frac{\lambda^{2}}{2}\,\langle\delta\hat{d}\,(\hat{K}+\omega)^{-1}\,\delta\hat{d}\rangle, \qquad
  \nc = \frac{\lambda^{2}\omega}{2}\,\langle\delta\hat{d}\,(\hat{K}+\omega)^{-2}\,\delta\hat{d}\rangle .
  \label{eq:exact}
\end{equation}
Thus, $m$ is real and nonnegative, and
$\dP = m - \nc = (\lambda^{2}/2)\langle\delta\hat{d}\,\hat{K}(\hat{K}+\omega)^{-2}\,\delta\hat{d}\rangle \ge 0$,
which is a second proof of Eq.~\eqref{eq:bound}.

At leading order in $\lambda$, the ground state becomes the molecular
ground state times a coherent state of the cavity mode, and $\hat{K}$
acting on $\delta\hat{d}|\Psi_0\rangle$ becomes
$\hat{H}_{\mathrm{el}} - E_{\mathrm{el}}$ of the isolated molecule. With the excitation
energies $\dE_k$ and transition dipoles $d_{0k} = \langle 0|\hat{d}|k\rangle$
of the isolated molecule, Eq.~\eqref{eq:exact} becomes
\begin{equation}
  m = \frac{\lambda^{2}}{2}\,\alpha^{+}(\omega), \qquad
  \nc = \frac{\lambda^{2}\omega}{2}\sum_{k}\frac{|d_{0k}|^{2}}{(\dE_k+\omega)^{2}},
  \label{eq:lo}
\end{equation}
where $\alpha^{+}(\omega) = \sum_k |d_{0k}|^{2}/(\dE_k+\omega)$ is the
antiresonant part of the dynamic polarizability along $\mathbf{e}$,
$\alpha(\omega) = \alpha^{+}(\omega) + \alpha^{+}(-\omega)$, and the same
$\alpha^{+}(\omega)$ enters the second-order energy of the coupled ground
state.\cite{Haugland2025} It follows from Eq.~\eqref{eq:lo} that
\begin{equation}
  \rsq \;=\; 1 + \frac{\Ebar(\omega)}{\omega}, \qquad
  \Ebar(\omega) = \frac{\sum_k \varpi_k\,\dE_k}{\sum_k \varpi_k}, \qquad
  \varpi_k = \frac{|d_{0k}|^{2}}{(\dE_k+\omega)^{2}},
  \label{eq:Ebar}
\end{equation}
so that $\rsq$ exceeds unity by a photon-weighted mean excitation energy
$\Ebar$ of the molecule measured in units of the cavity frequency.
By Eq.~\eqref{eq:exact}, this relation holds exactly with the excitation
energies and dipole matrix elements of the coupled system, which are
approximated by QED extensions of time-dependent density functional
theory\cite{Flick2019,Yang2021} and of configuration interaction singles.\cite{McTague2022,McTague2025}
Consistent with Eq.~\eqref{eq:bound}, the
squeezing,
$\dP = m - \nc = (\lambda^{2}/2)\sum_k |d_{0k}|^{2}\dE_k/(\dE_k+\omega)^{2}$,
is positive for any molecule. Unlike $\rsq$, it scales with the squared
transition dipoles, so that the two quantities can rank molecules
differently (Sec.~\ref{sec:survey}). Its $\omega \to 0$ limit,
$\lambda^{2}\alpha(0)/4$, is the squeezing obtained from the static
polarizability and exceeds $\dP$ at any finite cavity frequency.

The mean excitation energy is bounded by two ratios of the moments
$S_j = \sum_k f_k\,\dE_k^{\,j}$ of the oscillator strengths
$f_k = 2\dE_k|d_{0k}|^{2}$. Because $\Ebar(\omega)$ is non-decreasing in
$\omega$ (supplementary material), it lies between its limits
$\Ebar(0) = S_{-2}/S_{-3}$ and $\Ebar(\infty) = S_{0}/S_{-1}$, so that
\begin{equation}
  1 + \frac{S_{-2}}{S_{-3}\,\omega} \;\le\; \rsq \;\le\;
  1 + \frac{S_{0}}{S_{-1}\,\omega},
  \label{eq:bracket}
\end{equation}
where $S_{-2} = \alpha(0)$ is the static polarizability.

Therefore, at leading order, $\rsq$ and $\dP/\lambda^{2}$ are
linear-response properties of the isolated molecule (supplementary
material).
In contrast, the fourth moment requires quadratic response. At leading
order,
$\gtc = \rsq^{2} + R$, where $R$ follows from the quadratic response of the
molecule and describes photon pairs created together with an electronic
excitation.
$R$ equals 2 for harmonic matter but not in general (supplementary
material), so $\gtc$ is not determined by linear response.

\section{Computational Details}
\label{sec:comp}

Exact references were obtained using
full configuration interaction (FCI) electronic states combined
with a photon Fock space.\cite{Vu2024}
For H$_2$ and HeH$^+$, the Hamiltonian was
diagonalized directly. For LiH, the lowest root was obtained with a
matrix-free Davidson solver in the frame displaced by the ground-state
dipole of the isolated molecule, with at most five photons at
$\lambda = 0.02$~a.u. The $\lambda \to 0$ limits were evaluated
from FCI linear and quadratic response.
QED-CCSD-22 calculations\cite{Pavosevic2021,Pavosevic2022} used the
coherent-state QED-HF reference\cite{Haugland2020,Liebenthal2023} and the photon moments
of Ref.~\citenum{Brorsen2026}. Because the CC expectation values are
biorthogonal, $\langle\delta\hat{b}^{2}\rangle$ and
$\langle\delta\hat{b}^{\dagger 2}\rangle$ differ slightly, and we use their
mean for $m$, which is the combination that enters the CC value of $\Var(P)$.
Leading-order CC values were obtained by extrapolating
$\lambda = 0.005$, 0.01, and 0.02~a.u.\ to $\lambda \to 0$. All electrons
were correlated in the FCI and QED-CCSD-22 calculations.

Linear-response calculations used time-dependent Hartree--Fock (TDHF) and
time-dependent density functional theory\cite{Casida1995,Dreuw2005} with the
CAM-B3LYP,\cite{Yanai2004} PBE0,\cite{Adamo1999} and
B3LYP\cite{Becke1993,Stephens1994} functionals, without the Tamm--Dancoff
approximation. All excitations were obtained by full diagonalization of the
response matrix, and $\rsq$ was evaluated from Eq.~\eqref{eq:Ebar} as a sum
over all states. Calculations were performed with PySCF\cite{Sun2018,Sun2020}
using integration grid level 4 for the density functionals. The leading-order values of
Table~\ref{tab:lo} and Fig.~\ref{fig:omega} were obtained with the
aug-cc-pVTZ basis set and all other results with the aug-cc-pVDZ basis
set.\cite{Dunning1989,Kendall1992,WoonDunning1993,WoonDunning1994} For Li,
PySCF uses the earlier EMSL versions of these sets\cite{Feller1996,Schuchardt2007}
(supplementary material).

The first benchmark set contains five molecules, H$_2$, HeH$^+$, LiH, HF,
and H$_2$O, with the coupling vector along the bond or symmetry axis and, for
H$_2$O, also along $(1,1,1)/\sqrt{3}$, at $\omega = 5$, 10, and
20~eV (18 points). The second set contains eight molecules, N$_2$, CO, HCN,
C$_2$H$_2$, LiF, NH$_3$, CH$_4$, and CH$_2$O, with the coupling vector along
the principal axis and, for CH$_2$O, also along $(1,1,1)/\sqrt{3}$, at
$\omega = 2.5$, 5, and 10~eV (27 points). The references are exact
diagonalization for H$_2$ and HeH$^+$ and QED-CCSD-22 otherwise, all at
$\lambda = 0.02$~a.u. The geometries of H$_2$, HF, H$_2$O, NH$_3$, and
CH$_4$ are those of Ref.~\citenum{Weber2025}, and all geometries are given
in the Zenodo deposit.

For the survey, 29 molecules were optimized at the B3LYP/def2-SVP
level\cite{Weigend2005} starting from PubChem conformers,\cite{Kim2023,Bolton2011} and
their CAM-B3LYP/aug-cc-pVDZ spectra were computed with transition dipoles
along all three Cartesian axes. Values for an ensemble of molecules with fixed random orientations were obtained from the
orientational averages of $m$ and $\nc$ (supplementary material).

\section{Results and Discussion}
\label{sec:results}

\subsection{Weak-coupling theory against exact references}
\label{sec:validation}

Table~\ref{tab:lo} lists the leading-order photon statistics and mean
excitation energies in the aug-cc-pVTZ basis for the five molecules of
the first benchmark set, with the coupling vector along the bond or
symmetry axis.
In the aug-cc-pVDZ basis, FCI linear and quadratic response reproduce the
$\lambda \to 0$ limit of exact diagonalization, which confirms
Eqs.~\eqref{eq:lo} and \eqref{eq:Ebar} and the quadratic-response
expression for $R$ derived in the supplementary material. QED-CCSD-22
reproduces the FCI values to within a relative $6\times10^{-4}$
(supplementary material).
Because $\rsq - 1$ at $\lambda = 0.02$~a.u.\ differs from its
leading-order value by at most 0.21\% in this set, the references at this
coupling can be compared directly with leading-order linear response. The
change from aug-cc-pVDZ to aug-cc-pVTZ affects $\rsq - 1$ more than the
coupling does, but it changes the error of CAM-B3LYP relative to the
references by at most one percentage point (supplementary material).

\begin{table}
\caption{Leading-order photon statistics in the aug-cc-pVTZ basis: the
ratio $\rsq$ of Eq.~\eqref{eq:r}, the fourth-moment term $R$, the degree
of second-order coherence $\gtc = \rsq^{2}+R$, and the mean excitation
energy $\Ebar = (\rsq-1)\omega$, together with its sum-rule limits
$\Ebar(0)$--$\Ebar(\infty)$ of Eq.~\eqref{eq:bracket}. FCI linear and
quadratic response for H$_2$, HeH$^+$, and LiH; QED-CCSD-22 extrapolated to
$\lambda \to 0$ for HF and H$_2$O, whose sum-rule limits are from
CAM-B3LYP. Energies in eV.}
\label{tab:lo}
\begin{ruledtabular}
\begin{tabular}{lcccccc}
 & $\omega$ & $\rsq$ & $R$ & $\gtc$ & $\Ebar$ & $\Ebar(0)$--$\Ebar(\infty)$ \\
\hline
H$_2$ & 5 & 3.868 & 5.12 & 20.08 & 14.3 & 14.1--15.7 \\
 & 10 & 2.450 & 4.21 & 10.22 & 14.5 &  \\
 & 20 & 1.737 & 3.48 & 6.50 & 14.7 &  \\
HeH$^+$ & 5 & 6.559 & 4.43 & 47.46 & 27.8 & 27.6--32.9 \\
 & 10 & 3.799 & 3.93 & 18.36 & 28.0 &  \\
 & 20 & 2.416 & 3.41 & 9.25 & 28.3 &  \\
LiH & 5 & 2.249 & 7.69 & 12.75 & 6.2 & 5.0--13.4 \\
 & 10 & 1.694 & 5.27 & 8.14 & 6.9 &  \\
 & 20 & 1.389 & 3.96 & 5.89 & 7.8 &  \\
HF & 5 & 5.335 & 4.60 & 33.06 & 21.7 & 20.0--41.4 \\
 & 10 & 3.254 & 3.69 & 14.28 & 22.5 &  \\
 & 20 & 2.194 & 2.97 & 7.78 & 23.9 &  \\
H$_2$O & 5 & 4.604 & 4.89 & 26.09 & 18.0 & 16.0--33.6 \\
 & 10 & 2.906 & 3.79 & 12.23 & 19.1 &  \\
 & 20 & 2.024 & 3.01 & 7.10 & 20.5 &  \\

\end{tabular}
\end{ruledtabular}
\end{table}

\begin{figure}
\includegraphics[width=3.37in]{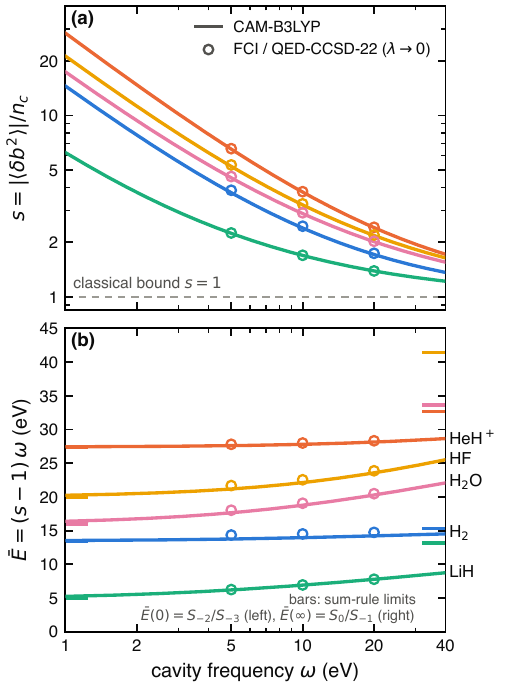}
\caption{(a) Ratio $\rsq$ of Eq.~\eqref{eq:r} and (b) mean excitation
energy $\Ebar = (\rsq-1)\omega$ as functions of the cavity frequency at
leading order (aug-cc-pVTZ). Lines are CAM-B3LYP linear response; open
circles are FCI (H$_2$, HeH$^+$, LiH) and QED-CCSD-22 extrapolated to
$\lambda \to 0$ (HF, H$_2$O). Bars in (b) mark the sum-rule limits
$\Ebar(0) = S_{-2}/S_{-3}$ (left) and $\Ebar(\infty) = S_0/S_{-1}$ (right)
from the CAM-B3LYP spectra (Table~\ref{tab:lo} gives the FCI limits for
H$_2$, HeH$^+$, and LiH).}
\label{fig:omega}
\end{figure}

Figure~\ref{fig:omega} shows $\rsq$ and $\Ebar$ as functions of the cavity
frequency. In all five systems, $\rsq$ exceeds unity at every frequency, as
required by Eq.~\eqref{eq:rge1}, and decreases toward unity as $\omega$
increases because $\Ebar$ varies much less than $\omega$. For H$_2$ and
HeH$^+$, $\Ebar$ stays close to its lower sum-rule limit, whereas for HF,
H$_2$O, and LiH it moves away from this limit as $\omega$ increases
(Fig.~\ref{fig:omega}b), reflecting the larger
spread of their oscillator strength over excitation energies. Finally, $R$
exceeds the harmonic value of 2 in every case (Table~\ref{tab:lo}).

\subsection{Linear-response route}
\label{sec:route}

\begin{figure}
\includegraphics[width=3.37in]{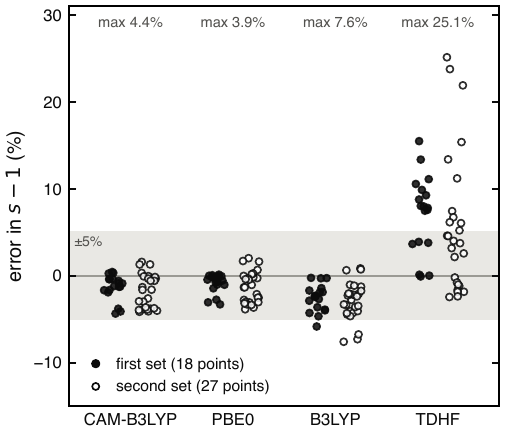}
\caption{Relative error in $\rsq - 1$ of linear-response calculations
against exact diagonalization (H$_2$, HeH$^+$) and
QED-CCSD-22 (all other systems), all in the aug-cc-pVDZ basis, at $\lambda = 0.02$~a.u. Filled circles:
first benchmark set (18 points, $\omega = 5$--20~eV); open circles: second
benchmark set (27 points, $\omega = 2.5$--10~eV). The shaded band marks
$\pm 5\%$.}
\label{fig:route}
\end{figure}

Figure~\ref{fig:route} and Table~\ref{tab:route} compare $\rsq$ from
linear-response calculations with the references at the 45 points of both
benchmark sets, which are 15 combinations of molecule and coupling
direction at three frequencies each. At leading order, this comparison
mainly tests how well each method describes the response of the isolated
molecule (Sec.~\ref{sec:weak}).
CAM-B3LYP and PBE0 reproduce both $\rsq - 1$ and the amount of
squeezing $\dP$ to within 5\% (Table~\ref{tab:route} and supplementary
material).
The errors are similar in both sets, including at the visible cavity
frequency of 2.5~eV.
On average, B3LYP underestimates $\rsq - 1$ in both sets
(Table~\ref{tab:route}), whereas TDHF overestimates it, because its
overestimated excitation energies raise $\Ebar$ directly through
Eq.~\eqref{eq:Ebar}.

\begin{table}
\caption{Signed mean, mean absolute, and maximum absolute relative errors
(\%) in $\rsq - 1$ of linear-response calculations against the references
of the first (18 points) and second (27 points) benchmark sets.}
\label{tab:route}
\begin{ruledtabular}
\begin{tabular}{lcccccc}
 & \multicolumn{3}{c}{First set} & \multicolumn{3}{c}{Second set} \\
 & Mean & MAE & Max & Mean & MAE & Max \\
\hline
CAM-B3LYP & $-1.3$ & 1.4 & 4.4 & $-1.7$ & 2.0 & 4.2 \\
PBE0 & $-0.9$ & 0.9 & 3.3 & $-1.3$ & 1.8 & 3.9 \\
B3LYP & $-2.6$ & 2.6 & 5.8 & $-3.0$ & 3.1 & 7.6 \\
TDHF & $+7.1$ & 7.2 & 15.5 & $+5.7$ & 6.7 & 25.1 \\

\end{tabular}
\end{ruledtabular}
\end{table}

The errors of CAM-B3LYP and PBE0 are comparable to the sensitivity of the
CC reference to the choice of biorthogonal estimate. In the second set,
using $\langle\delta\hat{b}^{\dagger 2}\rangle$ alone in place of the mean
raises the largest CAM-B3LYP deviation in $\rsq - 1$ to 5.0\% (supplementary material).
We therefore regard agreement within about 5\% as the resolution of this
comparison.

A harmonic model of the molecule built from its linear-response spectrum
gives $R = 2$. With the CAM-B3LYP spectrum, this model underestimates $\gtc$
most for LiH at 5~eV, by 47\% relative to QED-CCSD-22 at
$\lambda = 0.02$~a.u., so $R$ must be obtained from the quadratic response
of the molecule.

\subsection{Squeezing across chemical families}
\label{sec:survey}

\begin{figure*}
\includegraphics[width=\textwidth]{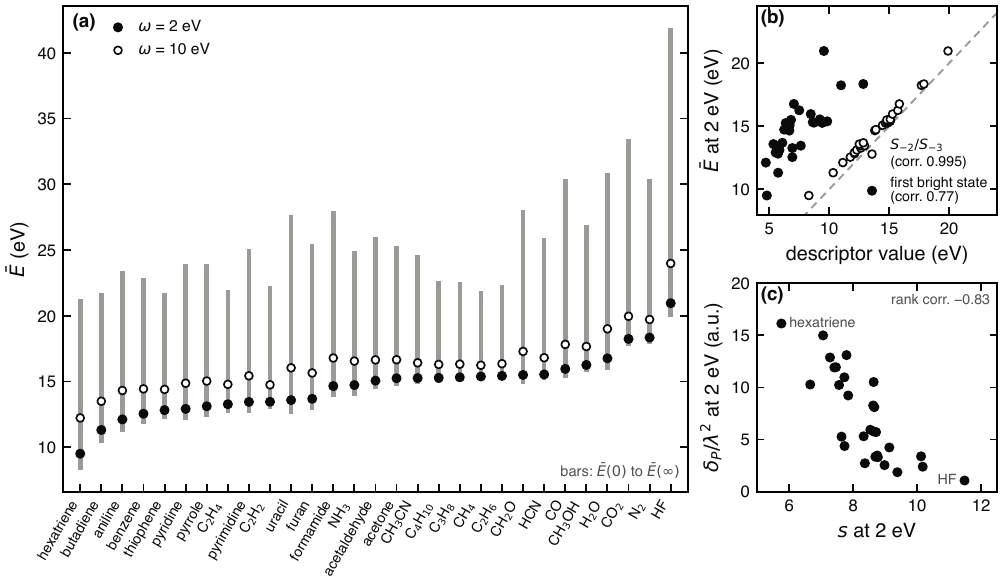}
\caption{(a) Mean excitation energy $\Ebar$ of 29 molecules for ensembles
with fixed random orientations at $\omega = 2$~eV (filled) and 10~eV (open),
CAM-B3LYP/aug-cc-pVDZ. Bars span the sum-rule limits $\Ebar(0)$ to
$\Ebar(\infty)$ of Eq.~\eqref{eq:bracket} for the same spectra. The finite
basis lowers the upper limits (supplementary material). (b) $\Ebar$ at 2~eV against
$S_{-2}/S_{-3}$ (open) and against the energy of the first bright
state, the lowest excited state with an isotropic oscillator strength
above 0.01 (filled). The dashed line is the diagonal. (c) Amount of
squeezing $\dP/\lambda^{2}$ against the ratio $\rsq$ at 2~eV for the same
ensembles.}
\label{fig:survey}
\end{figure*}

Figure~\ref{fig:survey} shows $\Ebar$ for the molecules of the survey
together with the sum-rule limits of Eq.~\eqref{eq:bracket}. At
$\omega = 2$~eV, a frequency typical of electronic strong-coupling
experiments, $\Ebar$ ranges from 9.5~eV for hexatriene to 21.0~eV for HF.
These values lie above their $\omega \to 0$ limit $S_{-2}/S_{-3}$, by much
less than the spread of $\Ebar$ across the molecules
(Fig.~\ref{fig:survey}b). The energy of the first bright state correlates
less closely with $\Ebar$ than $S_{-2}/S_{-3}$ does, so the squeezing ratio is governed by the ratio
of the static polarizability to the next inverse moment of the
oscillator-strength distribution rather than by the lowest bright state.

The amount of squeezing ranks the molecules differently. At 2~eV,
$\dP/\lambda^{2}$ is anticorrelated with $\rsq$ across the survey
(Fig.~\ref{fig:survey}c). Because $\dP/\lambda^{2}$ tends to $\alpha(0)/4$
as $\omega \to 0$ (Sec.~\ref{sec:weak}), it correlates strongly with the
static polarizability, with a correlation coefficient of 0.997. Thus, the
most polarizable molecules squeeze the field the most, although their
squeezing ratios are among the smallest.

Within chemical families, $\Ebar$ decreases with conjugation length, as
expected for a polarizability-weighted excitation energy (supplementary
material).

\subsection{Magnitude of the squeezing and polariton spectra}
\label{sec:magnitude}

For a single molecule, the squeezing is small. For water at
$\omega = 5$~eV and $\lambda = 0.05$~a.u., the squeezing
$-10\log_{10}[2\Var(P)]$ is only 0.03~dB. However, at fixed $\lambda$, $m$ and
$\nc$ of an ensemble of $N$ molecules both grow in proportion to $N$, while
$\rsq$ is unchanged at leading order, so the squeezing of an ensemble grows
with the collective coupling $\lambda\sqrt{N}$ (supplementary material).

\begin{figure}
\includegraphics[width=3.37in]{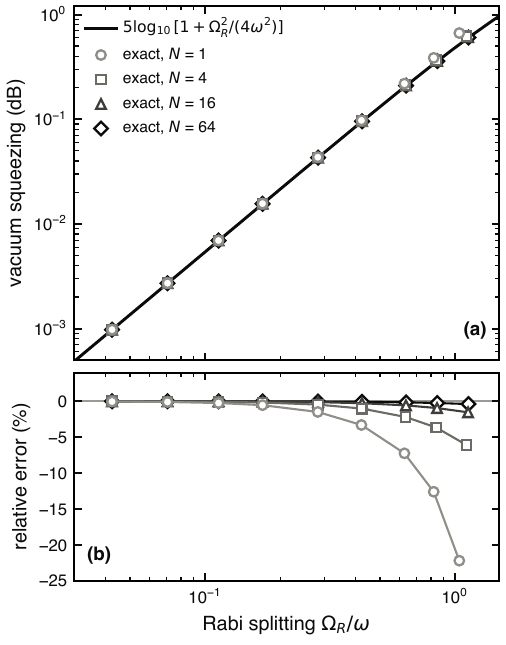}
\caption{Vacuum squeezing of the cavity field by $N$ two-level emitters in
resonance with the cavity as a function of the Rabi splitting $\Omega_{R}$
of the exact spectrum, in model units with unit excitation energy and
transition dipole and collective couplings $\lambda\sqrt{N} = 0.03$--0.8.
(a) Exact diagonalization (symbols) and the resonance form of
Eq.~\eqref{eq:twopol} (line). (b) Relative error of the resonance form.}
\label{fig:polariton}
\end{figure}

For an ensemble whose cavity spectrum consists of only a lower and an upper
polariton of a linearly polarized mode, Eq.~\eqref{eq:twopol} gives the
squeezing directly from the polariton frequencies, as for two coupled
oscillators.\cite{Gietka2024,Mirkhalaf2026}
For the Hopfield
model at resonance, $\omega_{+}\omega_{-} = \omega^{2}$, and
Eq.~\eqref{eq:twopol} reduces to a squeezing of
$5\log_{10}[1 + \Omega_{R}^{2}/(4\omega^{2})]$~dB, where
$\Omega_{R} = \omega_{+} - \omega_{-}$ is the Rabi splitting
(Fig.~\ref{fig:polariton}a).

Rabi splittings of 60--62\% of the transition energy have
been reached with a polymethine dye in an all-metal
microcavity.\cite{Barachati2018} For an ideal lossless single-mode cavity
with this normalized splitting, the resonance form gives about 0.2~dB.
Like the other decibel values in this section, this value refers to the
vacuum of the cavity mode as dressed by any nonresonant background in the
cavity. By the $\omega_{+} \to \infty$ limit of
Eq.~\eqref{eq:twopol}, a nondispersive background of permittivity
$\varepsilon$ alone lowers $\Var(P)$ below its empty-cavity value by a
factor $\varepsilon^{1/2}$, that is, by $5\log_{10}\varepsilon$~dB. In
decibels, the squeezing due to the
resonant coupling adds to this contribution (supplementary material).
Ground-state field fluctuations of this kind could be probed through the
Lamb shift of an ancilla qubit\cite{Lolli2015} or, for the transverse
electric field of the mode, by electro-optic sampling.\cite{DeLiberato2019}
This field is proportional to $\hat{X} - \lambda\hat{d}/\sqrt{\omega}$,
which is conjugate to $\hat{P}$, so it is anti-squeezed whenever $\hat{P}$
is squeezed.

Figure~\ref{fig:polariton} compares the resonance form with exact
diagonalization for $N$ two-level emitters in resonance with the cavity.
The error of the resonance form decreases with $N$ and remains small up to
$\Omega_{R} \approx \omega$ for the largest ensemble. For a single emitter,
the error grows with the splitting, because the product of the polariton
frequencies falls below $\omega^{2}$, whereas Eq.~\eqref{eq:twopol}
with the exact polariton frequencies remains within 3.1\%.

\subsection{Constraints for approximate methods}
\label{sec:constraints}

Equations~\eqref{eq:dEdw}--\eqref{eq:twopol} tie the ground-state squeezing
to the polariton spectrum and to the frequency dependence of the
ground-state energy, and they thus supply exact constraints on approximate
theories. The exact
ground-state energy satisfies Eq.~\eqref{eq:dEdw}, so that the exact
cavity-induced energy shift does not increase with $\omega$ at fixed
$\boldsymbol{\lambda}$, and approximate energy shifts should not increase
either. The identity itself holds for an approximate method when its
energy obeys the Hellmann--Feynman theorem with respect to $\omega$ and its
state obeys the virial relation and has $\langle\hat{P}\rangle = 0$, as the
exact ground state does.
Otherwise, the frequency derivative of its energy and its value of
$\Var(P)$ carry separate information (supplementary material).

QED-HF satisfies Eq.~\eqref{eq:dEdw} trivially, because in the coherent-state
basis\cite{Haugland2020,DePrinceYuwono2025,DePrinceYuwono2026} its energy does not depend on $\omega$ and its
photon field is a coherent state with $\Var(P) = 1/2$.
In contrast, the second-order energy used in the fluctuation--dissipation
functionals of Flick\cite{Flick2022} decreases with $\omega$ for a single mode
and any positive polarizability at imaginary frequency. With the polarizability of the
isolated molecule, its frequency derivative equals $-\dP$ of
Sec.~\ref{sec:weak}, so that it satisfies Eq.~\eqref{eq:dEdw} at leading
order in $\lambda$ (supplementary material).
Aklilu et al.\ also found that the ground-state energy of LiH obtained
with a QED density-functional method decreases as $\omega$ increases at
fixed $\boldsymbol{\lambda}$.\cite{Aklilu2026}
Finally, the QED-CCSD-22 energies of the five molecules of the first
benchmark set at $\lambda = 0.02$~a.u.\ decrease monotonically with $\omega$
between 5 and 20~eV, with slopes that agree with $-\dP$ from the CC photon
moments to within about 1\%, comparable to the difference between the two
biorthogonal estimates of $m$ (supplementary material).

Equation~\eqref{eq:rge1} sets the target for squeezed mean-field references,
such as the squeezed QED-HF reference of Mazin and Zhang\cite{Mazin2024} and
the squeezed strong-coupling QED-HF state proposed by El Moutaoukal et
al.,\cite{ElMoutaoukal2026} because the field of any coherent-state
reference satisfies $|m| \le \nc$. A reference whose field is a pure
squeezed state satisfies Eq.~\eqref{eq:rge1} automatically, because such a
state has $|m|^{2} = \nc(\nc + 1)$. Its ratio $\rsq = (1 + 1/\nc)^{1/2}$
then grows as $1/\lambda$ at weak coupling if $\nc$ has the value of
Eq.~\eqref{eq:lo}, whereas the exact ratio approaches $1 + \Ebar/\omega$.
Reproducing both moments therefore requires a field that is mixed by its
correlation with the electrons, which both references introduce through a
displacement that depends on the electronic configuration. Furthermore,
the displacements in both ans\"{a}tze shift only $\hat{X}$ and leave
$\Var(P)$ at its squeezed-vacuum value, so that Eq.~\eqref{eq:bound} rules
out a squeezing of $\hat{X}$ in the ground state. This condition
corresponds to $F \le 0$ in the notation of Mazin and Zhang and to
$\gamma \ge 0$ in that of El Moutaoukal et al.

The effective Hamiltonian that Tang et al.\ used to explain the squeezing
qualitatively\cite{Tang2025} gives the leading-order squeezing
$\lambda^{2}\alpha(0)/4$ at every cavity frequency. For every molecule of
the survey, this value exceeds $\dP$ by at least 20\% at 2~eV, so the
frequency dependence of the molecular response matters even at optical
cavity frequencies.

\section{Conclusions}
\label{sec:conclusions}

In this work, we used two exact sum rules for the field quadratures to
show that the vector-potential quadrature of a linearly polarized cavity
mode is never anti-squeezed in the ground state of any molecule at any
coupling strength within the dipole approximation. For the single-mode and
multimode Pauli--Fierz Hamiltonians in the dipole approximation, this result
proves the ground-state squeezing found in exactly solvable models and in
calculations for individual systems and is the nonperturbative counterpart
of the second-order results.

The bound
implies that the cavity field is squeezed
whenever the mode couples to a dipole-allowed transition from a
nondegenerate ground state, that the
cavity-induced shift of the ground-state energy cannot increase with the
cavity frequency at a fixed coupling vector, and that the incoherent photon
statistics are never sub-Poissonian. QED-HF meets the condition on the
energy shift trivially, whereas the second-order energy of
fluctuation--dissipation functionals decreases with the cavity frequency
for any positive polarizability, as do the QED-CCSD-22 energies examined
here.

At weak coupling, the squeezing ratio
exceeds unity by a photon-weighted mean excitation energy of the isolated
molecule in units of the cavity frequency, a relation that holds exactly
with the spectrum of the coupled system. Linear-response calculations with
the CAM-B3LYP and PBE0 functionals reproduce exact and QED-CCSD-22
references for both the squeezing and this ratio to within about 5\%.
Across 29 molecules, the mean excitation energy at a cavity frequency of
2~eV closely follows its $\omega \to 0$ limit, the ratio $S_{-2}/S_{-3}$ of
oscillator-strength moments, whereas the amount of squeezing is largest
for the most polarizable molecules.

Beyond leading order, the $\lambda^{2}$ coefficient of $\gtc$ from
QED-CCSD-22 is too large in magnitude by a factor of about 8 to 9 for LiH
(supplementary material), which is consistent with the missing connected
three-photon amplitudes. A QED-CC model with three-photon amplitudes is
therefore the natural next step for photon statistics at larger coupling
strengths. More broadly, the spectral form of
the bound connects the quantum fluctuations of the cavity vacuum to the
polariton spectrum and provides exact constraints for the development of
approximate light--matter methods.

\section*{Supplementary Material}

See the supplementary material for the proofs, derivations, and numerical
checks of the theoretical results, including the connection to the
photon-number expression of Ref.~\citenum{Weber2025}, the next-order theory,
QED-CCSD-22 beyond leading order, exact strong-coupling results, ensembles
of molecules, basis-set convergence, the benchmark and survey data, and a note
on gauges and quadrature labels.

\begin{acknowledgments}
This work was supported by the Air Force Office of Scientific Research under
AFOSR Award No.\ FA9550-24-1-0199. We acknowledge the use of artificial intelligence in assisting
writing some of the code used in this study and in editing and revising the manuscript.
\end{acknowledgments}

\section*{Author Declarations}

\subsection*{Conflict of Interest}

The author has no conflicts to disclose.

\subsection*{Author Contributions}

Kurt R. Brorsen: Conceptualization (lead), Software (lead), Writing (lead)

\section*{Data Availability}

The data that support the findings of this study are available within the
article and its supplementary material. The geometries, the
QED-CCSD-22/aug-cc-pVDZ energies of Sec.~S4, and the photon moments of the
benchmark sets are in the Zenodo deposit at
https://doi.org/10.5281/zenodo.23478204.
Additional data, including the
energies behind the basis-set and small-basis checks of Sec.~S4, are
available from the corresponding author upon reasonable request.

\bibliography{refs}

\clearpage
\newcount\SIpage
\SIpage=0
\loop
\advance\SIpage by 1
\noindent\thispagestyle{empty}\smash{\makebox[0pt][l]{\hspace*{\dimexpr-1in-\hoffset-\oddsidemargin\relax}\raisebox{\dimexpr-\paperheight+1in+\voffset+\topmargin+\headheight+\headsep+\topskip\relax}[0pt][0pt]{\includegraphics[page=\the\SIpage]{si.pdf}}}}
\par
\clearpage
\ifnum\SIpage<25
\repeat

\end{document}